\documentclass[preprint, pre,superscriptaddress] {revtex4}
\usepackage{color,graphics}
\usepackage{verbatim}
\usepackage{amsmath}
\usepackage{epsfig}

\usepackage{subfigure}

\newcommand{\lb}{{<}}
\newcommand{\rb}{{>}}
\newcommand{\ps}{pseudospinodal}

\newcommand{\ts}{\tau_s}
\newcommand{\tmax}{\tau_{\rm nequil}}
\newcommand{\txt}{{\tau_{\rm extrap}}} 
\newcommand{\tg}{\tau_g} 

\newcommand{\pb}{p_{\rm b}}
\newcommand{\ns}{n_s}

\newcommand{\sthres}{s^*}
\newcommand{\nuc}{nucleation}
\newcommand{\drop}{nucleating droplet}
\newcommand{\eq}{{\infty}}
\newcommand{\sech}{{\rm sech}}
\newcommand{\tobs}{t_{\rm obs}}
\newcommand{\rthres}{r^*}

\newcommand{\tmop}[1]{\ensuremath{\operatorname{#1}}}

\begin{document}

\title{Approaching equilibrium and the distribution of clusters}
\author{Hui Wang}
\affiliation{Department of Physics, Clark University, Worcester, MA 01610}
\author{Kipton Barros}
\affiliation{Department of Physics and the Center for Computational Science, Boston University, Boston, MA 02215}
\author{Harvey Gould}
\affiliation{Department of Physics, Clark University, Worcester, MA 01610}
\author{W.\ Klein}
\affiliation{Department of Physics and the Center for Computational Science, Boston University, Boston, MA 02215}

\begin{abstract}
We investigate the approach to stable and metastable equilibrium in Ising models using a cluster representation. The distribution of \nuc\ times is determined using the Metropolis algorithm and the corresponding $\phi^{4}$ model using Langevin dynamics. We find that the nucleation rate is suppressed at early times even after global variables such as the magnetization and energy have apparently reached their time independent values. The mean number of clusters whose size is comparable to the size of the nucleating droplet becomes time independent at about the same time that the nucleation rate reaches its constant value. We also find subtle structural differences between the \drop s formed before and after apparent metastable equilibrium has been established.
\end{abstract}

\maketitle

\section{Introduction}\label{sec:intro}

Understanding \nuc\ is important in fields as diverse as materials science,
biological physics, and meteorology~\cite{book,PorousSilicon, nanowires,
polydisperse, microtubule, ProteinNucleation, ProteinFolding, StaufferPRL82,bigklein}.
Fundamental progress was made when Gibbs assumed that the
\drop\ can be considered to be a fluctuation about metastable
equilibrium, and hence the probability of a 
\drop\ is independent of time~\cite{gunt}. 
Langer~\cite{lang} has shown that the probability of a \drop\
can be related to the analytic continuation of the stable state free energy
in the limit that the metastable state lifetime approaches infinity. Hence
the assumption by Gibbs is valid in this limit. It has also been shown that the
Gibbs assumption is correct in systems for which the interaction range $R \rightarrow
\infty$~\cite{uk,moreprecise}.

For metastable states with finite lifetimes equilibrium is never reached because 
a large enough fluctuation would initiate the transformation to the stable state. However, if the probability of 
such a fluctuation is sufficiently small, it is possible that systems investigated by simulations and experiments can be well 
approximated as being in equilibrium. Hence, for metastable lifetimes that are very long, we expect the 
Gibbs assumption to be a good approximation.

In practice, \nuc\ is not usually observed when the lifetime of the
metastable state is very long. Processes such as alloy formation, decay
of the false vacuum, and protein crystallization generally occur during a
continuous quench of a control parameter such as the temperature. It is
natural to ask if the nucleation process that is observed occurs when the
system can be reasonably approximated by one in metastable equilibrium. 
If so, the nucleation rate will be independent of time.

\begin{figure}[t] 
\begin{center}
\subfigure[\ $m(t)$.]{\label{fig: r_1_b_0_44_L_200_m_vs_t}\scalebox{0.7}{\includegraphics{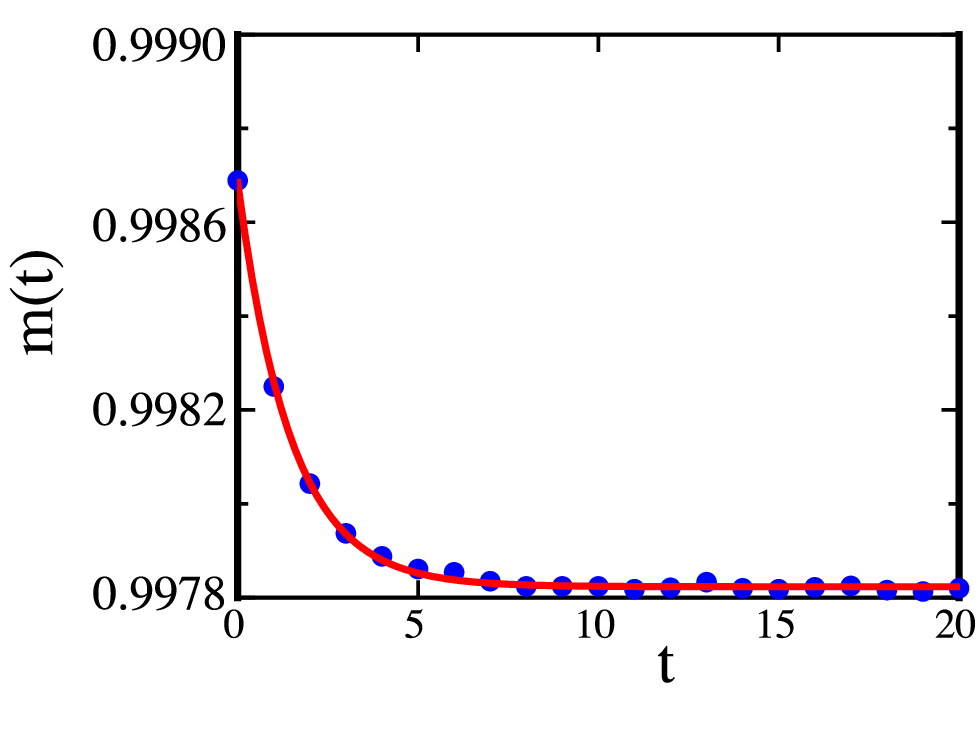}}}
\subfigure[\ $e(t)$.]{\label{fig: r_1_b_0_44_L_200_energy_vs_t}\scalebox{0.7}{\includegraphics{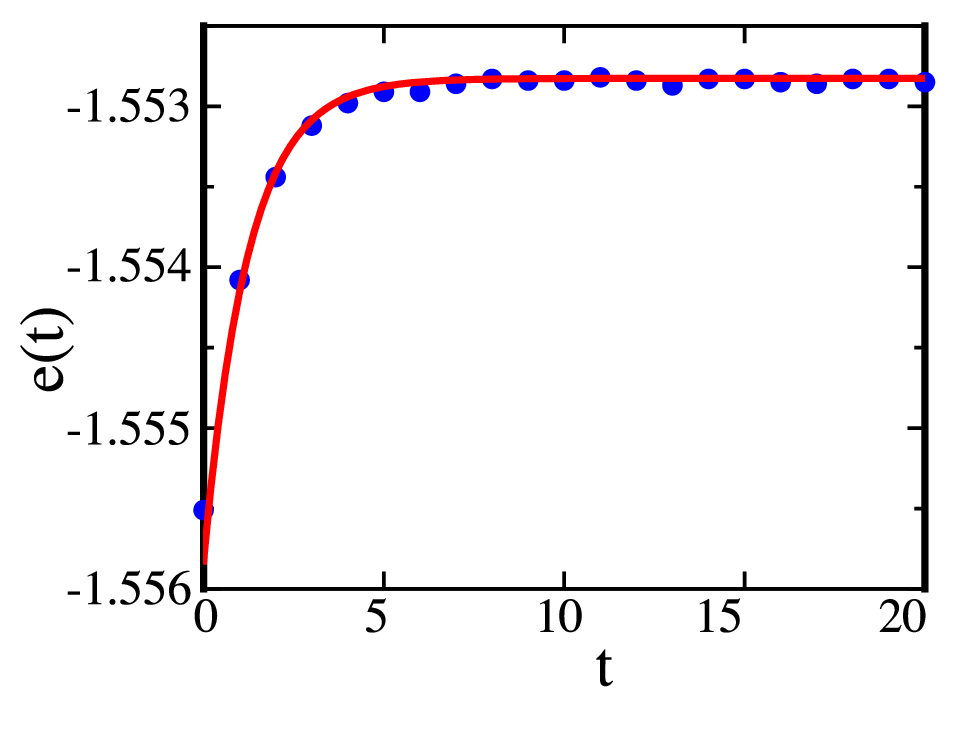}}}
\vspace{-0.1cm}
\caption{\label{fig:r_1_b_0_44_L_200_energy_acc_m}The evolution of the
magnetization $m(t)$ and the energy $e(t)$ per spin of the nearest-neighbor
Ising model on a square lattice with linear dimension $L = 200$ using the
Metropolis algorithm. The system was prepared at temperature $T=4T_c/9$ in the
external magnetic field $h = 0.44$. At time $t=0$ the sign of the magnetic field
is reversed. The solid lines are fits
to an exponential function with relaxation time $\tg = 1.5$ and 1.2 respectively. (Time is
measured in Monte Carlo steps per spin.) The data is averaged over 5000 runs.}
\end{center}
\end{figure}

It is usually assumed that metastable equilibrium
is a good approximation when the mean value of the order parameter and various
global quantities are no longer changing with time. As an example, we consider the nearest-neighbor Ising model on a square lattice and equilibrate the system at temperature $T=4T_c/9$ in a magnetic field $h = 0.44$. The relatively small value of the linear dimension $L=200$ was chosen in order to avoid \nuc\ occurring too quickly. At time $t=0$ the sign of the magnetic
field is reversed. In Fig.~\ref{fig:r_1_b_0_44_L_200_energy_acc_m} we plot the evolution of the magnetization $m(t)$ and the energy $e(t)$ per spin using the
Metropolis algorithm. The solid lines are the fits to an exponential function with the relaxation time $\tg \approx 1.5$. In the following we will
measure the time in terms of Monte Carlo steps per spin. A major goal of our work is to address the question, ``Can the system be treated as being in metastable equilibrium for $t \gtrsim \tg$?''

If the
nucleation rate is independent of time, the probability of 
a \drop\ occurring at time
$t$ after the change of magnetic field is an exponentially decreasing function of time. To
understand this dependence we divide the time into intervals $\Delta t$ and
write the probability that the system nucleates in a time interval $\Delta
t$ as $\lambda \Delta t$, where the \nuc\ rate $\lambda$ is a constant. The
probability that nucleation occurs in the time interval $(N + 1)$ is given by
\begin{equation}
\label{prob}
P_{N} = (1 - \lambda \Delta t)^{N}\lambda \Delta t.
\end{equation}
If we assume that $\lambda \Delta t$ is small and write $N=t/\Delta t$, we can write
\begin{align}
\label{probt}
P(t)\Delta t & = (1 - \lambda\Delta t)^{t/\Delta t}\lambda \Delta t \to e^{-\lambda t}\lambda \Delta t,
\end{align}
where $P(t) \Delta t$ is the probability that the system nucleates at a time between $t$ and $t + \Delta t$ after the change of the magnetic field. In the following we ask if the nucleation rate and the mean values of the order parameter and other thermodynamic quantities become independent of time at approximately the same time after a quench or is the approach to metastable equilibrium more complicated?

In Sec.~\ref{sec:distrib} we determine the probability distribution of the \nuc\ times and find that the \nuc\ rate becomes a constant only after a time $\tmax$ that is much longer than the relaxation time $\tg$ of $m(t)$ and $e(t)$. In Sec.~\ref{sec:relax} we study the microscopic behavior of the system and determine the relaxation time $\ts$ for $n_s$, the mean number of clusters of size $s$, to approach its equilibrium value~\cite{equil}. Our main result is that $\ts$ is an increasing function of $s$, and the time required for $n_s$ to reach its equilibrium value is the same order of magnitude as $\tmax$ for values of $s$ comparable to the \drop. That is, the time for the number of clusters that are the size of the \drop\ to reach its equilibrium value is considerably longer than the time for the mean value of the order parameter to become independent of time within the accuracy that we can determine.

In Secs.~\ref{sec:intervene} and \ref{sec:langevin} we show that there are subtle differences
between the structure of the \drop s which occur before and after metastable equilibrium appears to have been achieved. This difference suggests the possibility of finding even greater differences in the \drop s in systems of physical and technological importance. We summarize and discuss our results in Sec.~\ref{sec:summary}. In the Appendix we study the evolution of the clusters after a quench to the critical temperature of the Ising model and again find that that the clusters equilibrate in size order, with the smaller clusters equilibrating first. Hence in principle, an infinite system will never equilibrate. How close to equilibrium a system needs to be and on what spatial scale so that it can be treated by equilibrium methods 
depends on the physical process of interest.

\section{\label{sec:distrib}Distribution of \nuc\ times}

We simulate the Ising model on a square lattice with interaction range $R$ with the Hamiltonian
\begin{equation}
H = -J\!\sum_{\lb i,j \rb}s_{i}s_{j} - h\!\sum_i s_{i},
\end{equation}
where $h$ is the external field. The notation $\lb i,j\rb$ in the first sum
means that the distance between spins $i$ and $j$ is within the interaction range $R$. We studied
both nearest-neighbor ($R = 1$) and long-range interactions ($R \geq 20$). The
interaction strength $J$ is scaled as $J= 4/q$, where $q= 2R(R+1)$ is the number of
interaction neighbors per spin. The external field $h$ and the temperature are measured in terms of $J$. All of our simulations are at
temperature $T=4T_c/9$, where $T_{c}$ is the critical temperature. For
$R = 1$ the critical temperature is
$T_{c} \approx 2.269$. For $R \gtrsim 20$ the mean field result $T_{c} = 4$
is a good approximation to the exact value of the critical temperature~\cite{BinderPRE96}. As discussed in Sec.~\ref{sec:intro}
the system is equilibrated in a magnetic field $h$. The time $t = 0$ corresponds to the time
immediately after the magnetic field is reversed.

The clusters in the Ising model are defined rigorously by a mapping of the Ising critical point onto the
percolation transition of a properly chosen percolation model~\cite{lr_ising_cluster_def,ck,bigklein}. Two parallel spins that are within the interaction range $R$ are connected only if there is a bond between them. The bonds are assigned with the probability $\pb = 1 - e^{-2\beta J}$ for $R = 1$ and $\pb = 1 - e^{-2\beta J (1 - \rho)}$ near the spinodal, where $\rho$ is the density of the stable spins, and $\beta$ is the
inverse temperature. Spins that are connected by bonds form a cluster.

Because the intervention method~\cite{intervene} of identifying the \drop\ is time consuming (see Sec.~\ref{sec:intervene}), we use a simpler criterion in
this section to estimate the \nuc\ time. We monitor the size of the largest
cluster (averaged over 20 bond realizations) and estimate the \nuc\ time as the time when the largest cluster
first reaches a threshold size $\sthres$. The threshold size $\sthres$ is chosen
so that the largest cluster begins to grow rapidly once its size is greater than or equal to $\sthres$. Because $\sthres$ is larger than the actual
size of the \drop, the \nuc\ time that we estimate by this
criterion will be 1 to 2 Monte Carlo steps per spin later than the
\nuc\ time determined by the intervention method. Although the
distribution function $P(t)$ is shifted to slightly later times, the \nuc\ rate
is found to be insensitive to the choice of the threshold.

\begin{figure}[t]
\begin{center}
\subfigure[\ $P(t)$.]{\scalebox{0.75}{\includegraphics{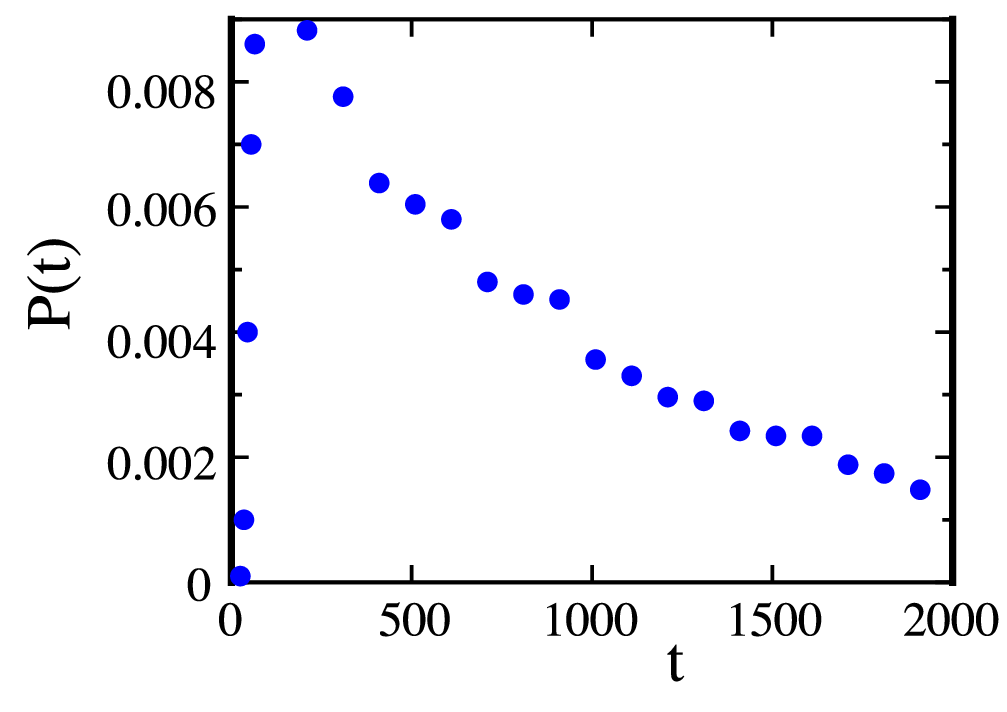}}}
\subfigure[\ $\ln P(t)$.]{\scalebox{0.75}{\includegraphics{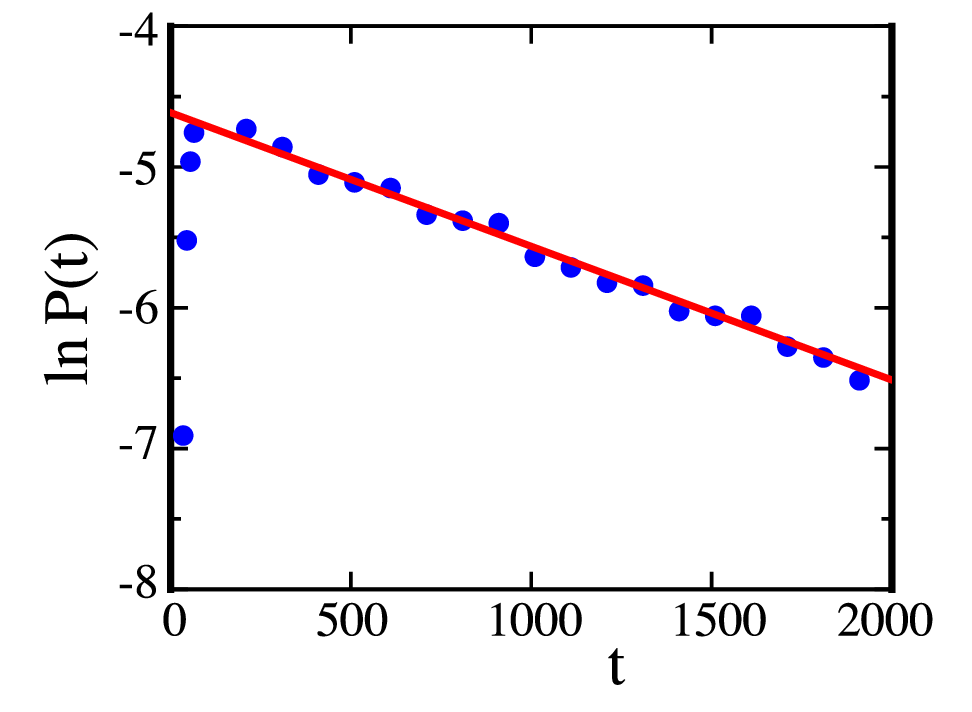}}}
\caption{\label{fig: hist_r_1_b_0_44}The distribution of \nuc\ times
$P(t)$ averaged over 5000 runs for the same system as in Fig.~\ref{fig:r_1_b_0_44_L_200_energy_acc_m}. The threshold size was chosen to be $\sthres = 30$. (The mean size of the \drop\ is $\approx 25$ spins.) (a) $P(t)$ begins to decay exponentially at $\tmax \approx
60$. The \nuc\ rate after equilibrium has been established is determined from the log-linear plot in (b) and is $\lambda \approx
9\times 10^{-4}$ (see Eq.~\eqref{probt}).}
\end{center}
\vspace{-0.25cm}
\end{figure}

Figure~\ref{fig: hist_r_1_b_0_44} shows $P(t)$ for $R=1$ and $h=0.44$, where $P(t)\Delta t$ is the probability that nucleation has
occurred between time $t$ and
$t+\Delta t$. The results for
$P(t)$ were averaged over 5000 runs. The mean
size of the \drop\ is estimated to be approximately 25 spins for this value of $h$. Note that $P(t)$ is an increasing
function of
$t$ for early times, reaches a maximum at $t=\tmax \approx 60$, and fits to
the expected exponential form for $t \gtrsim \tmax$. The fact that $P(t)$ falls below the expected exponential for $t < \tmax$ indicates that the
\nuc\ rate is reduced from its equilibrium value and that the system is not in metastable
equilibrium. Similar nonequilibrium effects have been observed in Ising-like~\cite{dieter, KBrendel} and continuous systems~\cite{Huitema}. We conclude that the
time for the
\nuc\ rate to become independent of the time after the change of magnetic field is much longer
than the relaxation time $\tg \simeq 1.5$ of the magnetization and energy. We will refer to nucleation that occurs before metastable equilibrium has been reached as {\it transient nucleation}.

\begin{figure}[t]
\begin{center}
\subfigure[\ $m(t)$.]{\label{fig:r_20_b_1_258_L_500_m_vs_t}\scalebox{0.8}{\includegraphics{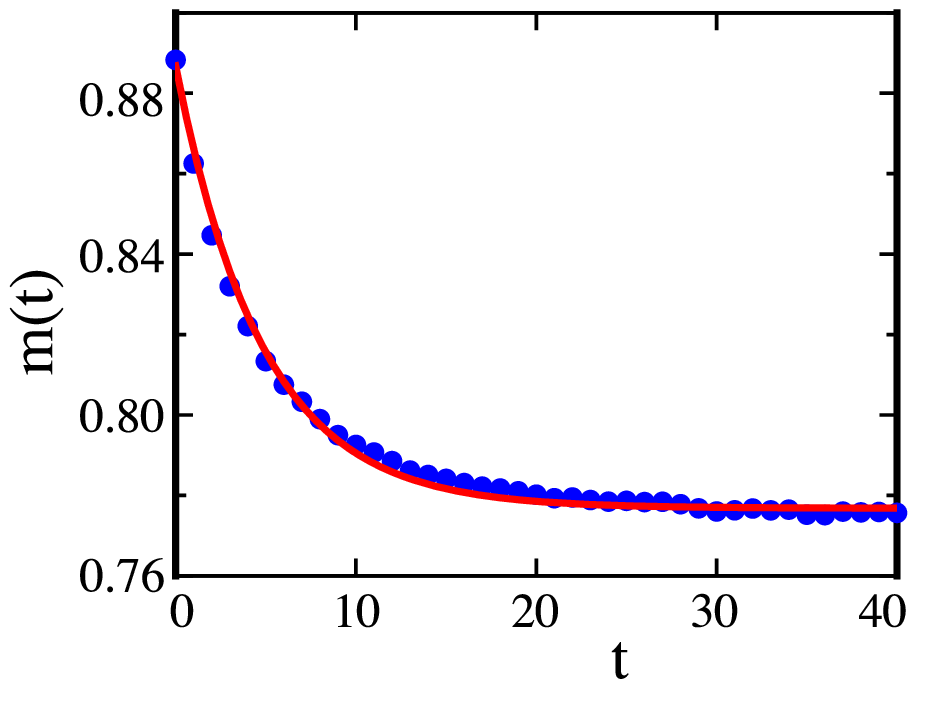}}}
\subfigure[\ $\ln(P(t))$.]{\label{fig: hist_r_20_b_1_258}\scalebox{0.8}{\includegraphics{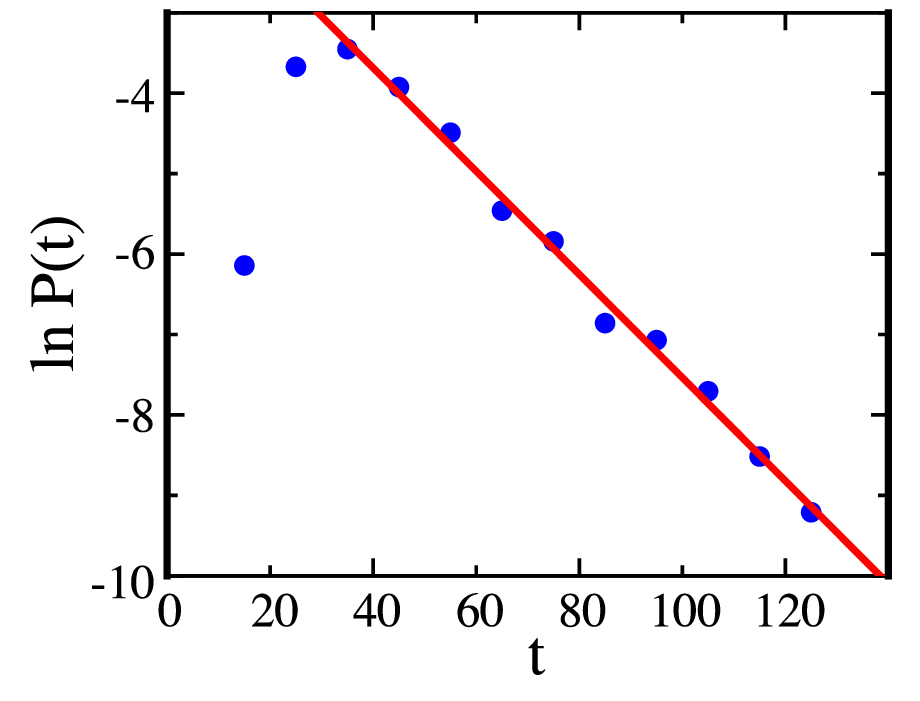}}}
\caption{\label{fig:r_20_b_1_258_L_500_energy_m} (a) The evolution of $m(t)$ for the long-range Ising model on a square lattice with $R = 20$, $h = 1.258$, and $L = 500$. The solid line is an exponential fit with the relaxation time $\tg \approx 4.5$. The data is averaged over 2000 runs. (b) The distribution of \nuc\ times $P(t)$ for the same system and number of runs. $P(t)$ decays exponentially for $t \gtrsim \tmax \approx
40$. The \nuc\ rate once equilibrium has been established is $\lambda = 6.4\times 10^{-2}$. The mean size of the nucleating droplet is $\approx 300$ spins.}
\end{center}
\vspace{-0.25cm}
\end{figure}

In order to see if the same qualitative behavior holds near the \ps, we
simulated the long-range Ising model with $R=20$ and $h=1.258$. In the
mean-field limit $R \to \infty$ the spinodal field is at $h_{\rm s}=1.2704$ (for $T=4T_c/9$).
A plot of $m(t)$ for this system is shown in Fig.~\ref{fig:r_20_b_1_258_L_500_m_vs_t} and is seen to have the same qualitative behavior as in Fig.~\ref{fig: hist_r_1_b_0_44} for $R=1$; the relaxation time $\tg \approx 4.5$. In Fig.~\ref{fig: hist_r_20_b_1_258} the distribution of nucleation times is shown, and we see that $P(t)$ does not decay exponentially until $t \gtrsim \tmax=40$. According to Ref.~\onlinecite{aaron}, $\tmax$ should become comparable to $\tg$ in the limit $R \to \infty$ because the free energy is described only by the magnetization in the mean-field limit. We find that the difference between $\tmax$ and $\tg$ is smaller for $R = 20$ than for $R = 1$, consistent with Ref.~\onlinecite{aaron}.

\section{\label{sec:relax}Relaxation of clusters to metastable equilibrium}

Given that there is a significant time delay between the relaxation of the
magnetization and the energy and the equilibration of the system as measured
by the \nuc\ rate, it is interesting to monitor the time-dependence of the cluster-size distribution after the
reverse of the magnetic field. After the change the system gradually relaxes to metastable
equilibrium by forming clusters of spins in the stable direction. How long is required for the number of clusters of size $s$ to reach equilibrium? In particular, we are interested in the time required for clusters that are comparable in size to the \drop.

\begin{figure}[b] 
\begin{center}
\scalebox{0.8}{\includegraphics{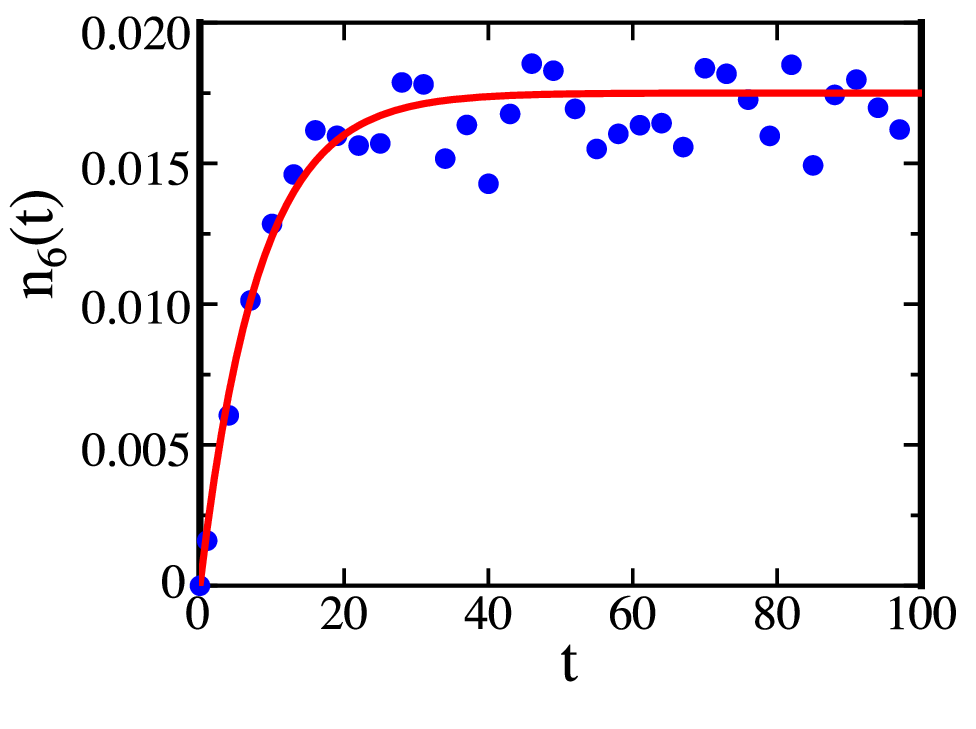}}
\vspace{-0.5cm}
\caption{\label{fig:r_1_b_0_44_L_200_ns_6_vs_t}The evolution of the number
of clusters of size $s = 6$ averaged over 5000 runs for $R = 1$ and the same
conditions as in Fig.~\ref{fig:r_1_b_0_44_L_200_energy_acc_m}. The fit is to the exponential form in Eq.~\eqref{eqn:exp} with $\ts \approx 8.1$ and $n_{s,\,\infty} =
0.0175$.}
\vspace{-0.25cm}
\end{center}
\end{figure}

\begin{figure}[t] 
\begin{center}
\subfigure[\ $R=1$.]{\label{fig5a}\scalebox{0.8}{\includegraphics{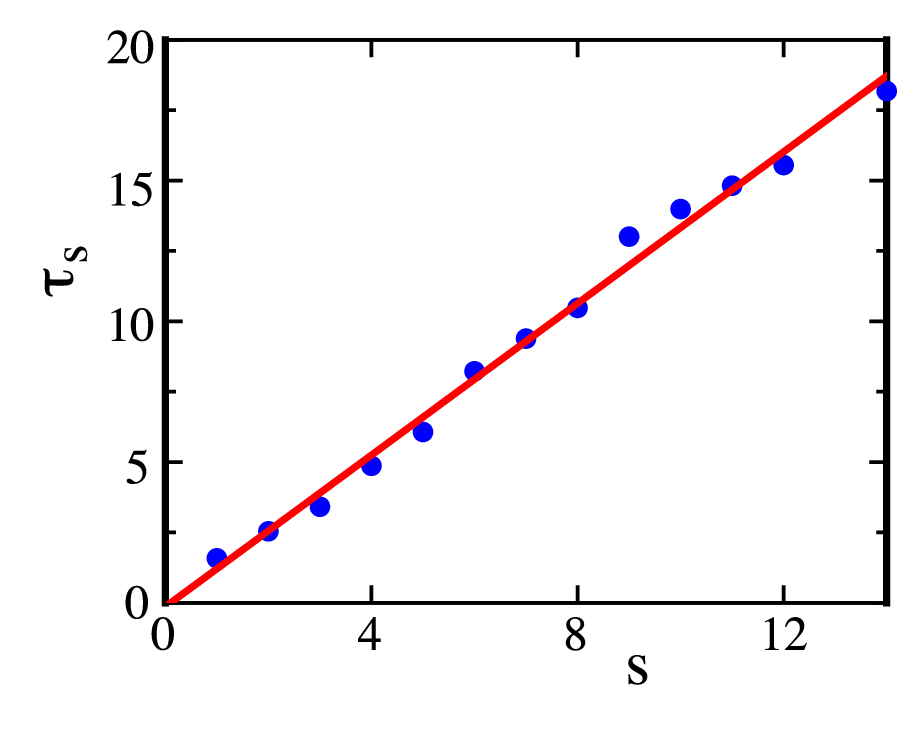}}}
\subfigure[\ $R=20$.]{\label{fig5b}\scalebox{0.8}{\includegraphics{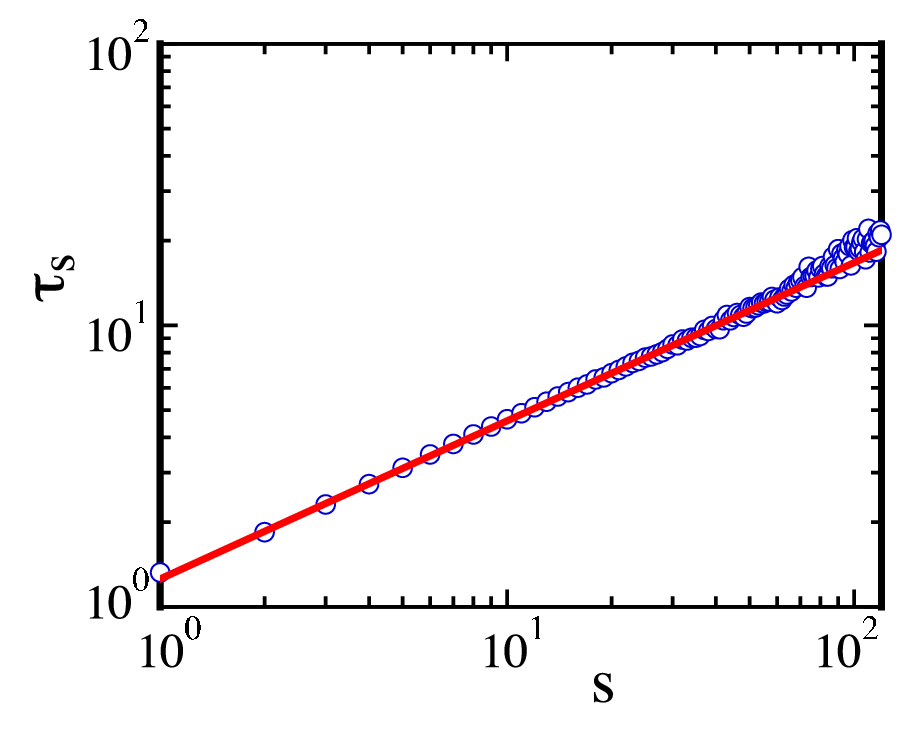}}}
\vspace{-0.1cm}
\caption{\label{fig: r_1_b_0_44_cluster_exp_fit_tau_s_metrop}(a) The equilibration time $\ts$ as a function of the cluster size $s$ for $R=1$ and $h=0.44$ the same conditions as in Fig.~\ref{fig:r_1_b_0_44_L_200_energy_acc_m}. The $s$-dependence of $\ts$ is approximately linear. The extrapolated value of $\ts$ corresponding to the
mean size of the \drop\ ($\approx 25$ spins) is $\txt \approx 34$, which is the same order of magnitude as time $\tmax \approx 60$ for the system to reach metastable equilibrium. (b) Log-log plot of the equilibration time $\ts$ versus $s$ for $R=20$ and $h=1.258$ and the same conditions as in Fig.~\ref{fig: hist_r_20_b_1_258}. We find that $\ts \sim s^x$ with the exponent $x\approx 0.56$. The extrapolated value of $\ts$ corresponding to the
mean size of the \drop\ ($\approx 300$ spins) is $\txt \approx 30$, which is comparable to the time $\tmax\approx 40$ for the system to reach metastable equilibrium.}
\vspace{-0.5cm}
\end{center}
\end{figure}

We first consider $R=1$ and monitor
the number of clusters $\ns$ of size $s$ at time $t$. To obtain good
statistics we chose $L = 200$ and averaged over 5000 runs.
Figure~\ref{fig:r_1_b_0_44_L_200_ns_6_vs_t} shows the evolution of
$n_{6}(t)$, which can be fitted to the exponential form:
\begin{equation}
\label{eqn:exp}
n_{s}(t) = n_{s,\,\infty}[1 - e^{-t/\ts}].
\end{equation}
We find that $\ts \approx 8.1$ for $s=6$. By doing similar fits for a range of $s$, we find that the time $\ts$ for the mean number of clusters of size $s$ to become time independent increases linearly with $s$ over the range of $s$ that we can simulate (see
Fig.~\ref{fig: r_1_b_0_44_cluster_exp_fit_tau_s_metrop}). The extrapolated value of $\ts$ corresponding to the
mean size of the \drop\ ($\approx 25$ spins by direct simulation) is $\txt \approx 34$. That is, it takes a time of $\txt \approx 34$ for the mean number of clusters whose size is the order
of the \drop s to become time independent. The time $\txt$ is much longer than the relaxation
time $\tg \approx 1.5$ of the macroscopic quantities $m(t)$ and $e(t)$ and is comparable to the time $\tmax \approx 60$ for the nucleation rate to become
independent of time.

Because the number of clusters in the \drop\ is relatively small for $R=1$ except very close to coexistence (small $h$), we also consider a long-range Ising model with $R = 20$ and $h = 1.258$ (as in Fig.~\ref{fig:r_20_b_1_258_L_500_energy_m}). The relaxation time $\ts$ of the clusters near the pseudospinodal fits to a power law $\ts \sim s^{x}$ with $x \approx 0.56$ (see Fig.~\ref{fig5b}). We know of no theoretical explanation for the qualitatively different dependence f the relaxation time $\ts$ on $s$ near coexistence ($\ts \simeq s$) and near the spinodal ($\ts \simeq s^{1/2}$). If we extrapolate $\ts$ to $s = 300$, the approximate size of the \drop, we find that the equilibration time for clusters of the size of the \drop\ is $\txt \approx 30$, which is comparable to the time $\tmax \approx 40$ for the nucleation rate to become independent of time.

To determine if our results are affected by finite size effects, we compared the equilibration time of the clusters for lattices with linear dimension $L = 2000$ and $L = 5000$. The equilibration times of the clusters were found to be unaffected.

\section{\label{sec:intervene}Structure of the \drop}

Because \nuc\ can occur both before and after the system is in metastable equilibrium, we ask if there are any structural differences between the \drop s formed in these two cases. To answer this question, we determine the nature of the \drop s for the one-dimensional (1D) Ising model where we can make $R$ (and hence the size of the \drop s) large enough so that the structure of the \drop s is well defined. In the following we take $R = 2^{12}= 4096$, $h = 1.265$, and $L = 2^{18}$. The relaxation time for $m(t)$ is $\tg \approx 40$, and the time for the distribution of nucleation times to reach equilibrium is $\tmax \approx 90$.

We use the
intervention method to identify nucleation~\cite{intervene}. To
implement this method, we choose a time at
which a \drop\ might exist and make many copies of
the system. Each copy is restarted using a
different random number seed. The idea is to
determine if the largest cluster in each of the copies
grows in approximately the same place at about the
same time. If the percentage
of copies that grow is greater than 50\%, the
\drop\ is already in the growth phase; if it is
less than 50\%, the time chosen is earlier than
\nuc. We used a total of 20 trials to make this determination.

Our procedure is to observe the system
for a time $\tobs$ after the intervention
and determine if the size of the largest cluster
exceeds the threshold size $\sthres$ at
approximately the same location. To ensure that
the largest cluster at $\tobs$ is the same
cluster as the original one, we require that the
center of mass of the largest cluster be within a
distance $\rthres$ of the largest cluster in
the original configuration. If these conditions
are satisfied, the \drop\ is said to grow. We 
choose $\tobs = 6$, $\rthres
= 2R$, and $\sthres = 2000$. (In comparison,
the size of the \drop\ for the particular run that
we will discuss is $\approx 1080$ spins.)

There is some ambiguity in our identification of the \nuc\ time because the saddle point parameter is large but finite~\cite{bigklein}. This ambiguity manifests itself in the somewhat arbitrary choices of the parameters $\tobs$, $\rthres$, and $\sthres$. We tried
different values for $\tobs$, $\rthres$, and $\sthres$ and found that our
results depend more strongly on the value of the parameter $\rthres$
than on the values of $\tobs$ and $\sthres$. If we take $\rthres = R/2$, the \drop s almost always occur one to two Monte Carlo steps per
spin later than for $\rthres = 2R$. The reason is that the linear
size of the \drop\ is typically 6 to $8R$, and its center of mass might
shift more than $R/2$ during the time $\tobs$. If such a shift
occurs, a cluster that would be said to grow for $\rthres = 2R$ would not be counted as such because it did not
satisfy the center of mass criterion. This shift causes an
overestimate of the time of the \drop. A reasonable choice of $\rthres$ is 20\% to 40\% of the linear size of the \drop. The choice of
parameters is particularly important here because the rate of growth of the transient
\drop s is slower than the growth rate of droplets formed after
metastable equilibrium has been reached. Hence, we have to identify the
nucleating droplet as carefully as possible.

Because nucleation studies are computationally
intensive, we used a novel algorithm for
simulating Ising models with a uniform long-range
interaction ~\cite{kip}. The algorithm uses a
hierarchical data structure to store the 
magnetization at many length scales, and can find
the energy cost of flipping a spin in time $O((\ln
R)^d)$, rather than the usual time $O(R^d)$, where
$d$ is the spatial dimension.

\begin{figure}[b]
\begin{center}
\scalebox{1.0}{\includegraphics{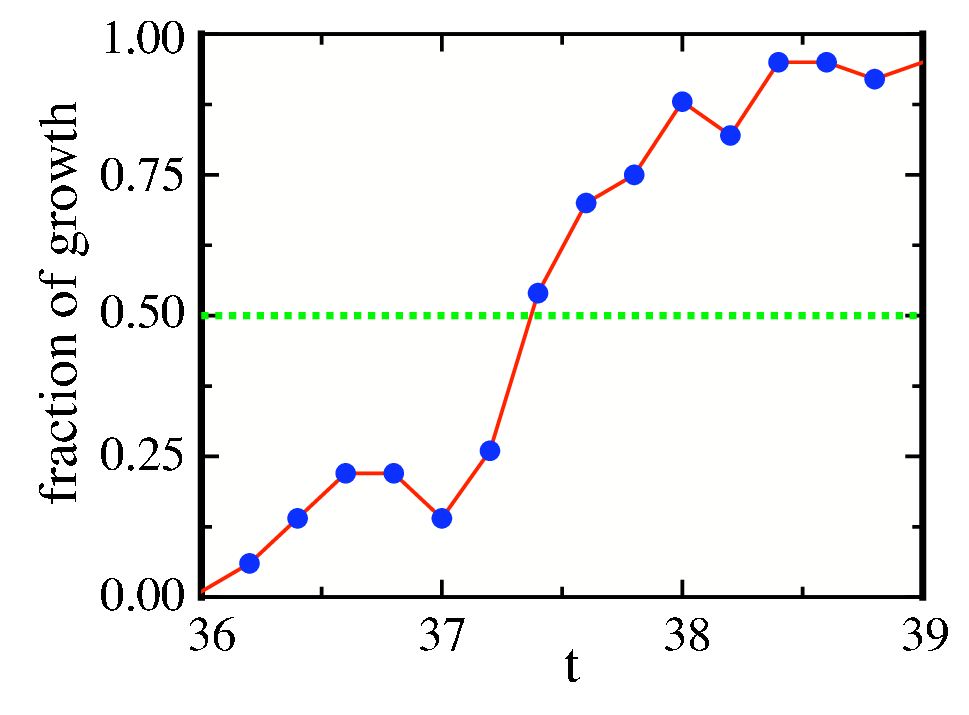}}
\vspace{-0.5cm}
\caption{\label{fig:growth_fraction_for_intervention}The fraction of copies for which the largest cluster grows for a particular run for a 1D Ising model with $R = 2^{12}$, $h = 1.265$, and $L = 2^{18}$. The time for 50\% growth is $\approx 37.4$. The largest cluster at this time corresponds to the \drop\ and has $\approx 1080$ spins. For this intervention 100 copies were considered; twenty copies were considered for all other runs.}
\end{center}
\end{figure}

Figure~\ref{fig:growth_fraction_for_intervention} shows the fraction of copies for which the largest cluster grows as a function of the intervention time. For this particular run the \drop\ is found to occur at $t\approx 37.4$.

\begin{figure}[t]
\begin{center}
\subfigure[\ Comparison to Eq.~\eqref{eq:profile}.]{\label{fig7a}\scalebox{0.8}{\includegraphics{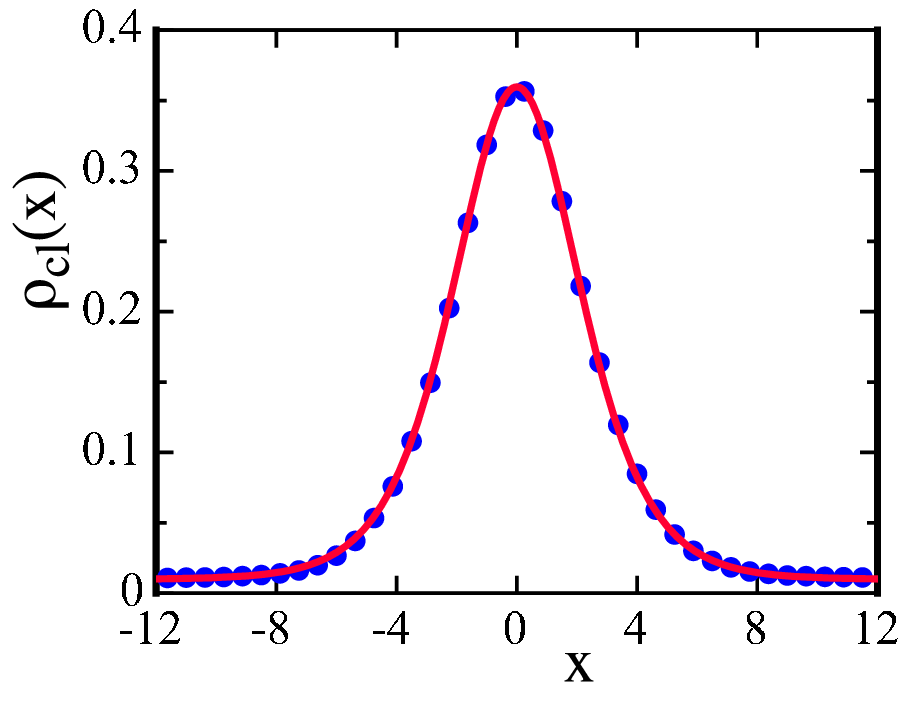}}}
\subfigure[\ Comparison to Gaussian.]{\label{fig7b}\scalebox{0.8}{\includegraphics{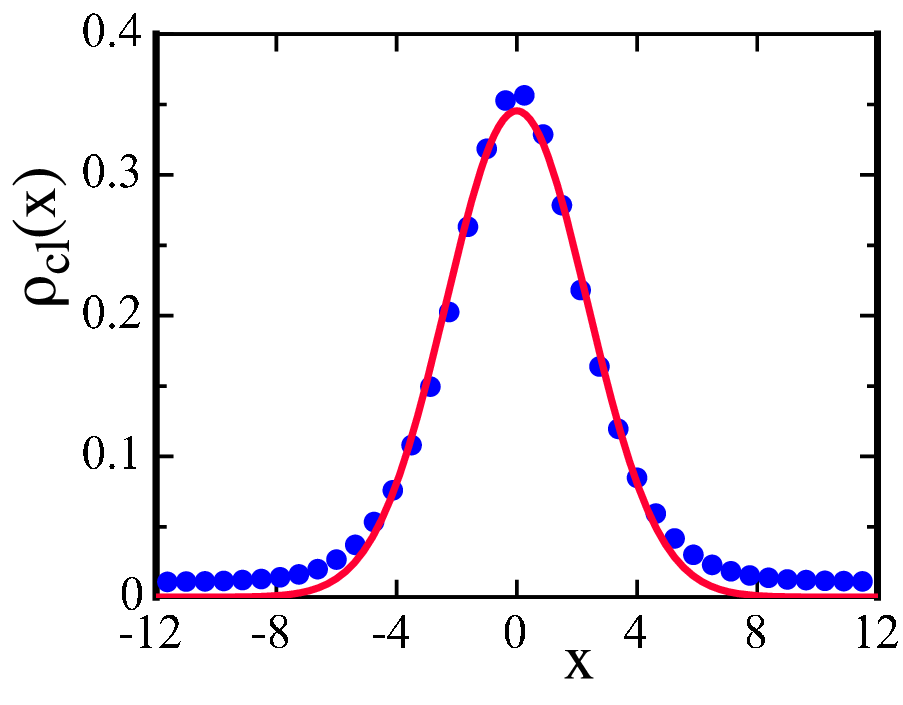}}}
\caption{\label{fig:equilprofile}Comparison of the mean cluster profile ($\bullet$) in the 1D Ising model after metastable equilibrium has been established with (a) the form in Eq.~\eqref{eq:profile} and (b) a Gaussian. Note that Eq.~\eqref{eq:profile} gives a better fit than the Gaussian, which underestimates the peak at $x=0$ and the wings. The $x$ axis is measured in units of $R$.}
\end{center}
\vspace{-0.25cm}
\end{figure}

We simulated 100 systems in which nucleation occurred before global quantities such as $m(t)$ became independent of time, $t < \tg \approx 40$, and 100 systems for which nucleation occurred after the nucleation rate became time independent ($t > \tmax \approx 90$). We found that the mean size of the \drop\ for $t < \tg $ is $\approx 1200$ with a standard deviation of $\sigma \approx 150$ in comparison to the mean size of the \drop\ for $t>\tmax$ of $\approx 1270$ and $\sigma \approx 200$. That is, the \drop s formed before metastable equilibrium has been reached are somewhat smaller.

We introduce the {\it cluster profile} $\rho_{\rm cl}$ to characterize the shape of the largest cluster at the time of nucleation. For a particular bond realization a spin that is in the stable direction might or might not be a part the largest cluster due to the probabilistic nature of the bonds. For this reason bond averaging is implemented by placing 100 independent sets of bonds between spins with probability $\pb = 1 - e^{-2\beta J (1 - \rho)}$ in the stable direction. The clusters are identified for each set of bonds, and the probability $p_i$ that spin $i$ is in the largest cluster is determined. The values of $p_i$ for the spins in a particular bin are then averaged using a bin width equal to $R/4$. This mean value of $p_{i}$ is associated with $\rho_{\rm cl}$. Note that the spins that point in the unstable direction are omitted in this procedure. The mean cluster profile is found by translating the peak position of each droplet to the origin.

Figure~\ref{fig7a} shows the mean cluster profile formed after metastable equilibrium has been established ($t >\tmax \approx 90$). The position $x$ is measured in units of $R$. For comparison we fit $\rho_{\rm cl}$ to the form
\begin{equation}
\label{eq:profile}
\rho(x) = A\,\sech^{2}(x/w) + \rho_0,
\end{equation}
with $A_{\rm cl}=0.36$, $w_{\rm cl}=2.95$ and $\rho_0=0$ by construction. In Fig.~\ref{fig7b} we show a comparison of $\rho_{\rm cl}$ to the Gaussian form $A_g\exp(-(x/w_g)^2)$
with
$A_g = 0.35$ and $w_g= 3.31$. Note that Eq.~\eqref{eq:profile} gives a better fit than a Gaussian, which underestimates the peak at $x=0$ and the wings. Although Unger and Klein~\cite{uk} derived Eq.~\eqref{eq:profile} for the magnetization saddle point profile, we see that this form also provides a good description of the cluster profile. 

\begin{figure}[t]
\begin{center}
\scalebox{1.0}{\includegraphics{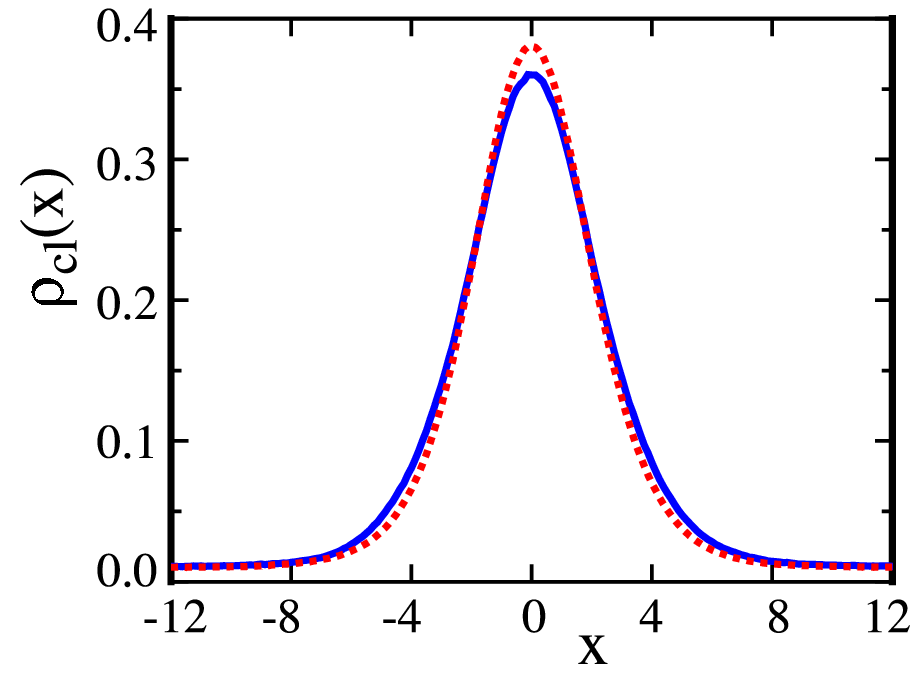}}
\vspace{-0.25cm}
\caption{\label{fig:compare}The cluster profiles of the \drop s formed before (dashed line) and after (solid line) metastable equilibrium has been established. Both profiles are consistent with the form given in Eq.~\eqref{eq:profile}, but the transient \drop s are slightly more compact. The fitting parameters are $A = 0.38$ and $w = 2.67$ for the transient droplets and $A = 0.35$ and $w = 2.95$ for the droplets formed after the nucleation rate has become independent of time.}
\end{center}
\vspace{-0.25cm}
\end{figure}

A comparison of the cluster profiles formed before and after metastable equilibrium is shown in Fig.~\ref{fig:compare}. Although both profiles are consistent with the form in Eq.~\eqref{eq:profile},
the transient \drop s are more compact, in agreement with the predictions in Ref.~\onlinecite{aaron}.

\begin{figure}[t]
\begin{center}
\subfigure[\ Comparison with Eq.~\eqref{eq:profile}.]{\label{fig9a}\scalebox{0.8}{\includegraphics{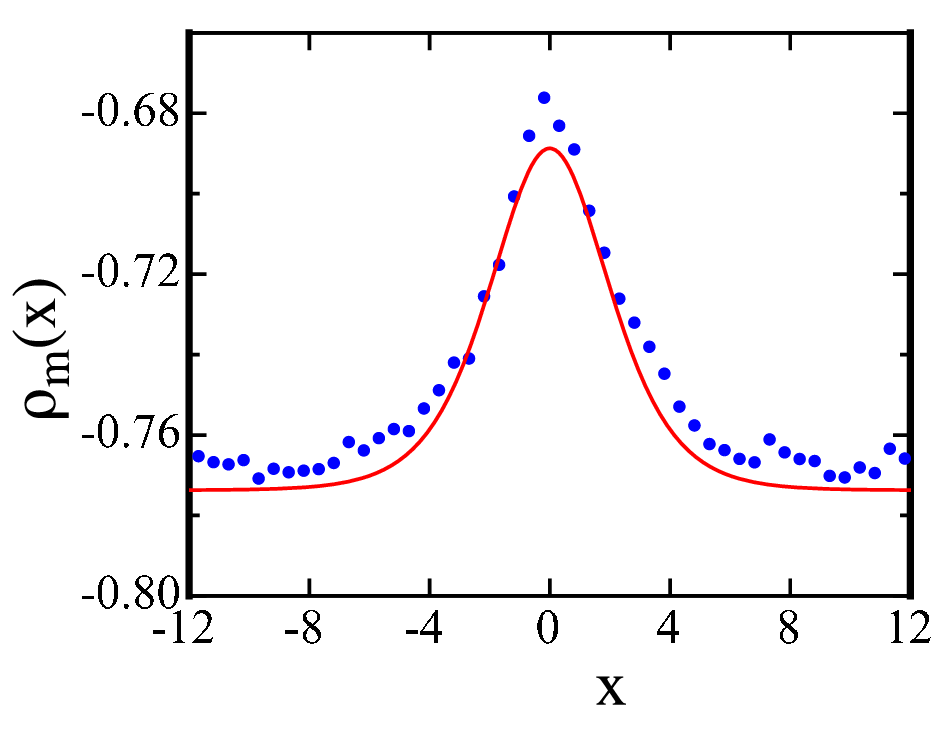}}}
\subfigure[\ Comparison of profiles.]{\label{fig9b}\scalebox{0.8}{\includegraphics{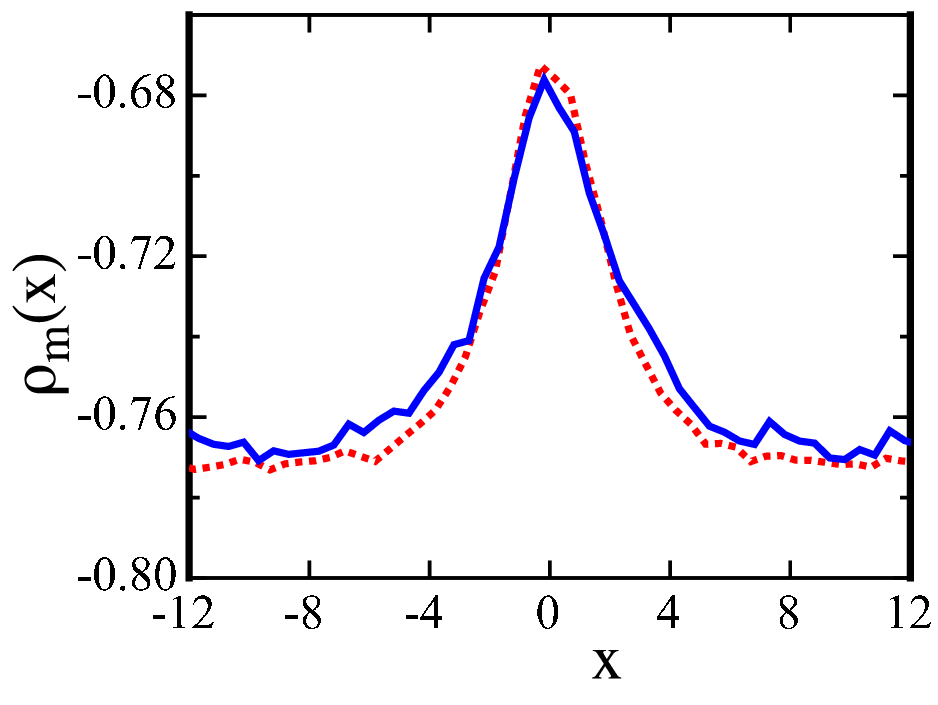}}}
\vspace{-0.05cm}
\caption{\label{fig:compare_profile_m}(a) The magnetization density profile of the \drop s formed after metastable equilibrium has been established. The solid line is the analytical solution~\cite{ising_analytical} which has the form in Eq.~\eqref{eq:profile} with the calculated values $A = 0.085$, $w = 2.65$, and $\rho_0 = -0.774$. (b) Comparison of the density profile of \drop s formed before (dashed line) and after (solid line) metastable equilibrium has been established by coarse graining the magnetization. The same qualitative differences between the \drop s that occur before and after metastable equilibrium are observed as in Fig.~\ref{fig:compare}, although the magnetization density profile is much noisier than the cluster density profile.}
\end{center}
\vspace{-0.15cm}
\end{figure}

We also directly coarse grained the spins at the time of nucleation to obtain the density profile of the coarse-grained magnetization $\rho_{\rm m}(x)$ (see Fig.~\ref{fig9a}). The agreement between the simulation and analytical results~\cite{ising_analytical} are impressive, especially considering that the analytical form is valid only in the limit $R \to \infty$. The same qualitative differences between the \drop s that occur before and after metastable equilibrium is found (see Fig.~\ref{fig9b}), although the magnetization density profile is much noisier than that based on the cluster analysis.

\section{\label{sec:langevin}Langevin simulations}

It is interesting to compare the results for the Ising model and the Langevin dynamics of the $\phi^4$ model. 
One advantage of studying the Langevin dynamics of the $\phi^4$ theory is that it enables the efficient simulation of systems with a very large interaction range $R$. If all lengths are scaled by a large value of $R$, the effective magnitude of the noise decreases, making faster simulations possible.

The coarse grained Hamiltonian analogous to the 1D ferromagnetic Ising model with long-range interactions in an external field $h$ can be expressed as
\begin{equation}
\label{eq:H}
H[\phi] = - \frac{1}{2} \Big(R \frac{d \phi}{d x} \Big)^2 + \epsilon \phi^2 + u \phi^4 - h \phi,
\end{equation}
where $\phi(x)$ is the coarse-grained magnetization. A dynamics consistent with this Hamiltonian is given by,
\begin{equation}
\label{dynamics}
\frac{\partial \phi}{\partial t} = - M
\frac{\delta H}{\delta \phi} + \eta = - M \big[\!- R^2 \frac{d^2 \phi}{dx^2} + 2
\varepsilon \phi + 4 u \phi^3 - h \big] + \eta,
\end{equation}
where $M$ is the mobility and $\eta (x, t)$ represents zero-mean 
Gaussian noise with 
$\langle \eta (x, t) \eta (x', t')\rangle = 2 kT M \delta (x - x') \delta (t - t')$.

For nucleation near the spinodal the potential $V = \varepsilon \phi^2 +
u \phi^4 - h \phi$ has a metastable well only for $\varepsilon <
0$. The magnitude of $\phi$ and $h$ at the spinodal are given by
$h_{\tmop{s}} =\sqrt{(8|\varepsilon|^3/27 u)}$ and
$\phi_{\tmop{s}} = \sqrt{(|\varepsilon|/6 u)}$,
and are found by setting $V' = V'' = 0$. The distance from the spinodal is characterized by
the parameter
$\Delta h = |h_{\tmop{s}} - h|$.
For $\Delta h/h_{\tmop{s}} \ll 1$, the bottom of the metastable well $\phi_{\min}$ is
near $\phi_{\tmop{s}}$, specifically
$\phi_{\min} = -\phi_{\tmop{s}} (1 + \sqrt{2 \Delta h/3
h_{\tmop{s}}})$.

The stationary solutions of the dynamics are found by setting $\delta H/\delta
\phi = 0$. Besides the two uniform solutions corresponding to the minima in $V$,
there is a single nonuniform solution which approximates the nucleating droplet profile when the nucleation barrier is large. When
$\Delta h/h_{\tmop{s}} \ll 1$, the profile of the \drop\ is described by Eq.~\eqref{eq:profile} with $A = \sqrt{h_{\tmop{s}}/6\Delta h}/\phi_{\tmop{s}}$, $w = (8h_{\tmop{s}}\Delta h \phi_{\tmop{s}}^{2}/3)^{-1/4}$, and $\rho_0=\phi_{\min}$~\cite{uk}.

The dynamics (\ref{dynamics}) is numerically integrated using the scheme~\cite{scheme}
\begin{equation}
\label{discrete} \phi (t + \Delta t) = \phi (t) - \Delta tM \big[- R^2
\frac{d^2 \phi}{dx^2} + 2 \varepsilon \phi + 4 u \phi^3 - h\big] +
\sqrt{\frac{\Delta t}{\Delta x}} \eta,
\end{equation}
where $d^2 \phi/dx^2$ is replaced by its central difference
approximation. Numerical stability requires that $\Delta t < (\Delta x/R)^2$, but it is
often desirable to choose $\Delta t$ even smaller for accuracy.

As for the Ising simulations, we first prepare an equilibrated system with
$\phi$ in the stable well corresponding to the direction of the external field $h$. At $t =
0$ the external field is reversed so that the system relaxes to metastable
equilibrium. We choose $M = 1$, $T = 1$,
$\varepsilon = - 1$, $u = 1$, and $\Delta h = 0.005$. The scaled length of the system is chosen to be $L/R = 300$. We choose $R$ to be large
so that, on length scales of $R$, the metastable $\phi$ fluctuates near its equilibrium value $\phi_{\min} \approx
-0.44$.

After nucleation occurs $\phi$
will rapidly grow toward the stable well. To determine the distribution of nucleation times, we assume that when the value of the field $\phi$ in any bin reaches $0$, nucleation has
occurred. This relatively crude criterion is sufficient for determining the distribution of nucleation times if we assume
that the time difference between the nucleation event and its later detection
takes a consistent value between runs.

\begin{figure}[t]
\begin{center}
\scalebox{0.95}{\includegraphics{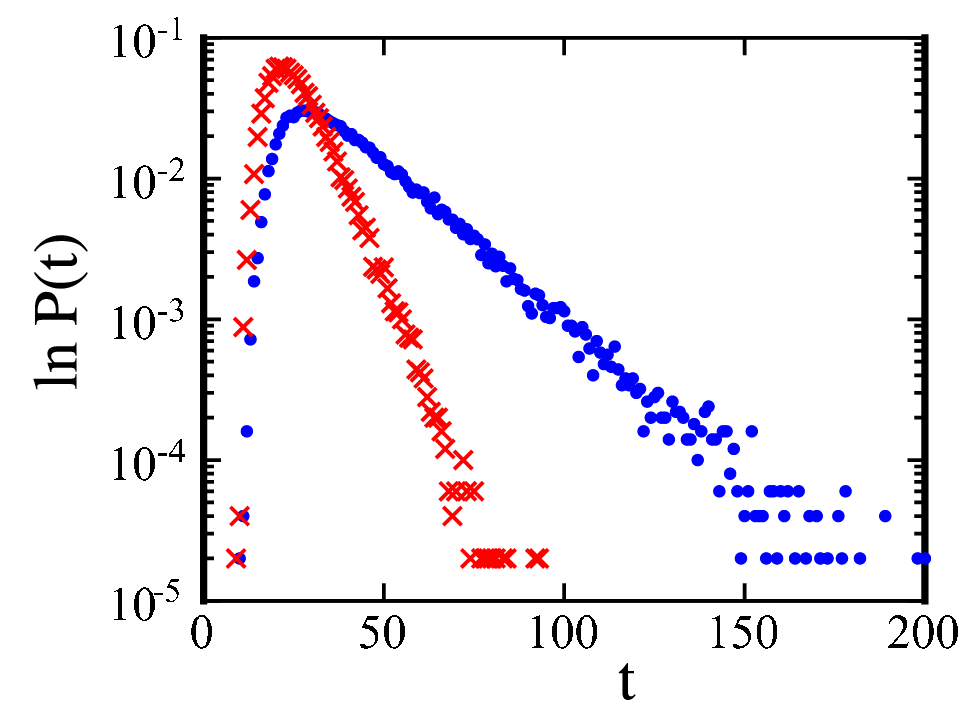}}
\vspace{-0.6cm}
\caption{\label{fig:langevin_nuc_dist2}Log-linear plot of the distribution $P(t)$ of \nuc\ times for the one-dimensional Langevin equation with $R=2000$ ($\times$) and $R = 2500$ ($\bullet$) averaged over 50,000 runs. The distribution is not exponential for early times, indicating that the system is not in metastable equilibrium. Note that the nucleation rate is a rapidly decreasing function of $R$.}
\vspace{-0.1cm}
\end{center}
\end{figure}

Figure~\ref{fig:langevin_nuc_dist2} compares the distribution of
50,000 nucleation times for systems with $R = 2000$ and $R = 2500$ with $\Delta x/R = 1$ and $\Delta t = 0.1$. The
distribution shows the same qualitative behavior as found in the Metropolis
simulations of the Ising model (see Fig.~\ref{fig: hist_r_1_b_0_44}). For example, the distribution of
nucleation times is not exponential for early times after the quench. As expected,
the nucleation rate decreases as $R$ increases. Smaller values of $\Delta x$ and $\Delta t$ give similar results for the distribution.

To find the droplet profiles, we need to identify the time of
nucleation more precisely. The intervention criterion, which was applied in Sec.~\ref{sec:intervene}, is one
possible method. In the Langevin context we can employ a simpler criterion: nucleation is considered to have occurred if $\phi$ decays to the saddle-point profile (given by Eq.~\eqref{eq:profile} for $\Delta h/h_{\tmop{s}} \ll 1$) when $\phi$ is evolved using noiseless dynamics~\cite{roy,aaron}. For fixed $\Delta h$ these two criteria agree in the $R \to \infty$ limit,
but can give different results for finite $R$~\cite{explain_critiria}.

\begin{figure}[t]
\begin{center}
\scalebox{0.95}{\includegraphics{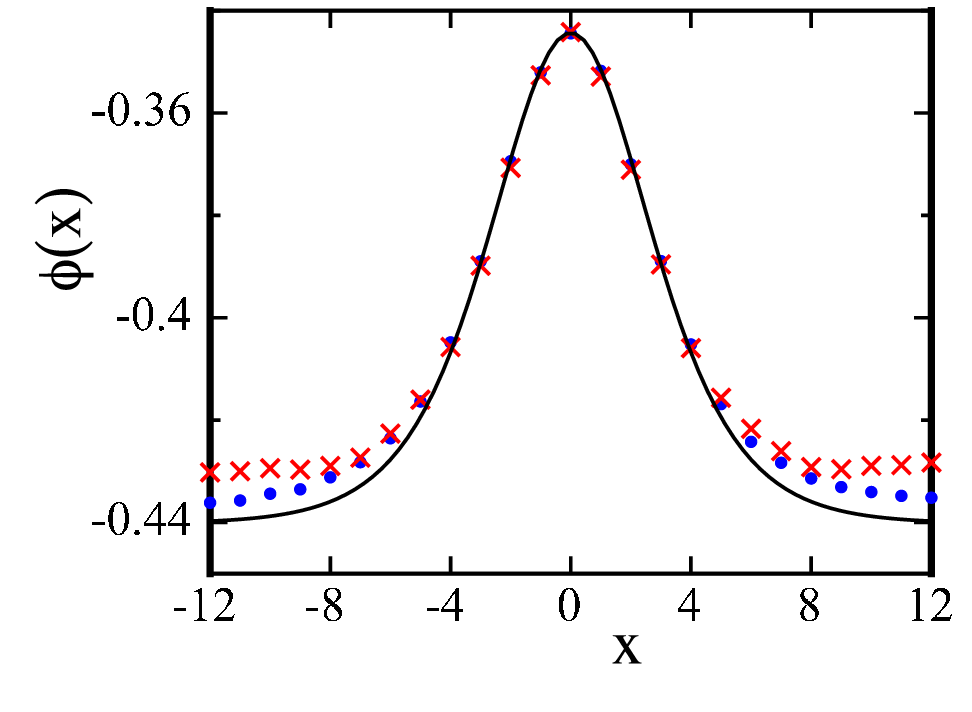}}
\vspace{-0.5cm}
\caption{\label{fig:langevin_profile1}Comparison of the density profile $\phi(x)$ of the nucleating droplets found by numerically solving the Langevin equation after metastable equilibrium has been reached for $R=2000$ ($\times$) and $R=4000$ ($\bullet$) to the theoretical prediction (solid line) from Eq.~\eqref{eq:profile} using the calculated values $A= 0.096$, $w= 3.58$, and $\rho_0 = -0.44$. The numerical solutions are averaged over 1000 profiles. The results suggest that as $R$ increases, the observed nucleation profiles converge to the prediction of mean-field theory.}
\vspace{-0.25cm}
\end{center}
\end{figure}

In Fig.~\ref{fig:langevin_profile1} we plot the average of 1,000 density profiles of the nucleating
droplets formed after metastable equilibrium has been established for $R=2000$ and $R=4000$. Note that there are noticeable deviations of
the averaged profiles from the theoretical prediction in Eq.~\eqref{eq:profile}, but the deviation is less for $R=4000$. The deviation is due to the fact that the
bottom of the free energy well in the metastable state is skewed; a similar deviation was also observed in the Ising model. We also note that the
individual nucleating droplets look much different
from their average. It is
expected that as $R$ increases, the profiles of the individual nucleating droplets will
converge to the form given by Eq.~\eqref{eq:profile}.

\begin{figure}[h!]
\begin{center}
\scalebox{0.95}{\includegraphics{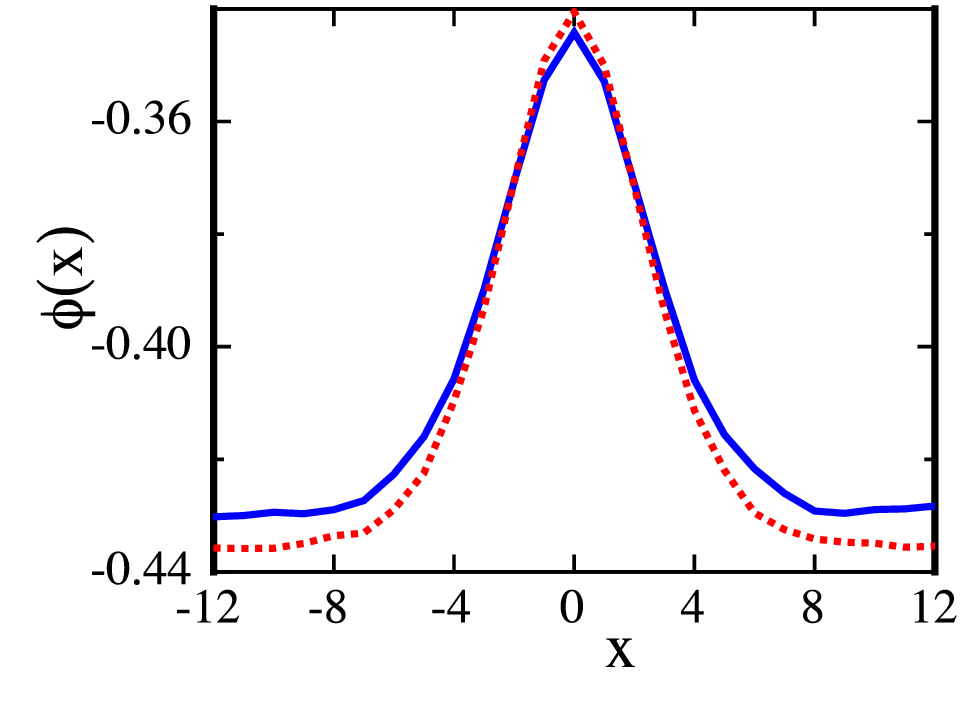}}
\vspace{-0.7cm}
\caption{\label{fig:langevin_profile2}The density profile of the nucleating droplets found from numerical solutions of the Langevin equation formed before (dotted line) and after (solid line) metastable equilibrium has been established. Nucleation events occurring before $t=15$ are transient, and events occurring for $t \geq 30$ are metastable. Both plots are the result of 1000 averaged profiles with an interaction range $R=2000$.}
\vspace{-0.5cm}
\end{center}
\end{figure}

In Fig.~\ref{fig:langevin_profile2} we compare the
average of 1,000 density profiles of nucleating droplets before and after metastable equilibrium has been established. As for the Ising model, there are subtle differences consistent with the predictions of Ref.~\onlinecite{aaron}. The transient
droplets have slightly lower background magnetization and compensate by being
denser and more compact.

\section{Summary}\label{sec:summary}

Although the time-independence of the mean values of macroscopic quantities such as the magnetization and the energy is often used as an indicator of metastable equilibrium, we find that the observed relaxation time of the clusters is much longer for sizes comparable to the \drop. This longer relaxation time explains the measured non-constant nucleation rate even when global quantities such as the magnetization appear to be stationary. By identifying the \drop s in the one-dimensional long-range Ising model and the Langevin equation, we find structural differences between the \drop s which occur before and after metastable equilibrium has been reached. Our results suggest that using global quantities as indicators for metastable equilibrium may not be appropriate in general, and distinguishing between equilibrium and transient nucleation is important in studying the structure of \drop s. Further studies of transient nucleation in continuous models of more realistic systems would be of interesting and practical importance.

Finally, we note a subtle implication of our results. For a system to be truly in equilibrium would require that the mean number of clusters of all sizes  be independent of time. The larger the cluster, 
the longer the time that would be required for the mean number to become time independent. Hence, the bigger the system, the longer the time that would be required for the system to reach equilibrium. Given that the system is never truly in metastable equilibrium so that the ideas of Gibbs, Langer, and others are never exactly applicable, when is the system close enough to equilibrium so that any possible simulation or experiment cannot detect the difference? We have found that the magnetization and energy are not sufficient indicators for nucleation and that the answer depends on the process being studied. 
For nucleation the equilibration of the number of clusters whose size is comparable to the size of the nucleating droplet is the relevant indicator.

\appendix

\section{Relaxation of clusters at the critical temperature}

Accurate determinations of the dynamical critical exponent $z$ have been found from the relaxation of the magnetization and energy at the critical
temperature. In the following we take a closer look at the relaxation of the
Ising model by studying the approach to equilibrium of the distribution of
clusters of various sizes.

We consider the Ising model on a square
lattice with $L = 5000$. The system is initially equilibrated at either zero
temperature $T_{0} = 0$ (all spins up) or at $T_{0} = \infty$, and then
instantaneously quenched to the critical temperature $T_c$. The
Metropolis algorithm is used.

\begin{figure}[t] 
\begin{center}
\subfigure[\ $R = 1$.]{\scalebox{0.8}{\includegraphics{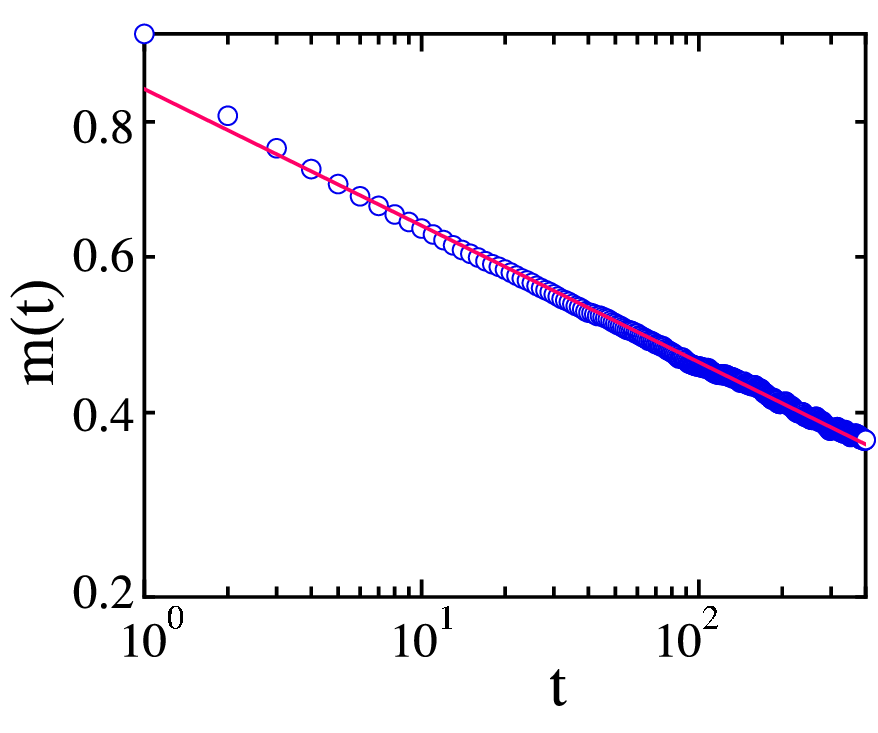}}}
\subfigure[\ $R = 128$.]{\scalebox{0.8}{\includegraphics{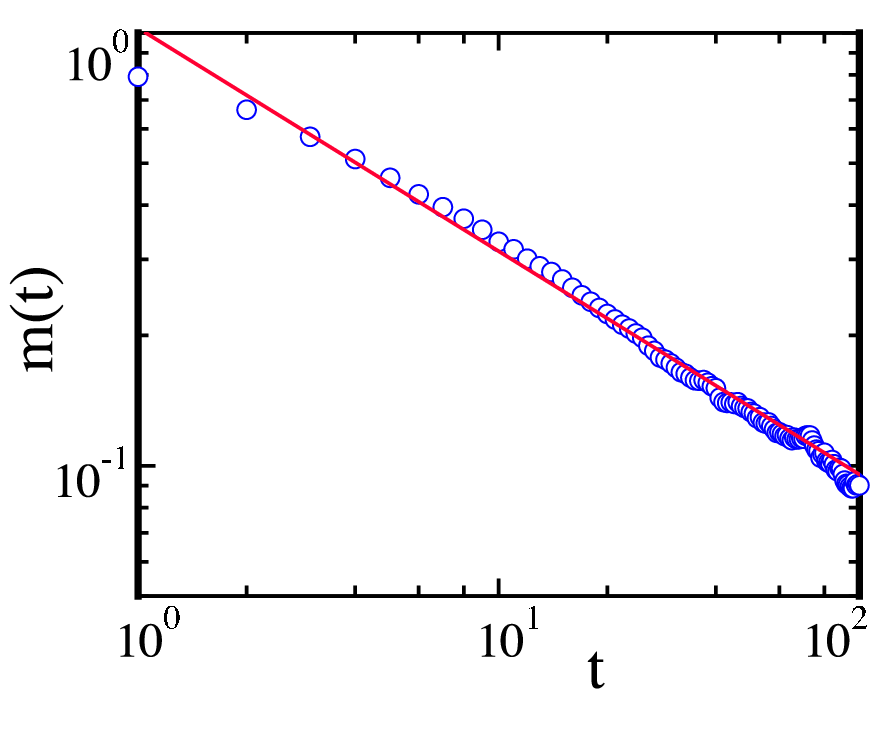}}}
\caption{\label{fig: metrop_m_t_Tc_up_log} The relaxation of the magnetization $m(t)$ of the 2D Ising
model at $T=T_c$ starting from $T_0=0$. (a) $R=1$, $T_c = 2.269$, $L = 5000$. (b) $R = 128$, $T_c = 4$, $L = 1024$. The straight line is the fit to a power law with
slope $\approx 0.057$ for $R=1$ and slope $\approx 0.51$ for $R=128$.}
\end{center}
\end{figure}

As a check on our results we first determine $m(t)$ starting from $T_0=0$. Scaling arguments suggest that $m(t)$ approaches its equilibrium value as~\cite{dieter2}
\begin{equation}
\label{eqn:power_eqn}
f(t) = Bt^{-\beta/\nu z} + f_{\eq},
\end{equation}
where the static critical exponents are $\beta = 1/8$ and $\nu = 1$ for finite $R$ and $\beta = 1/2$ and $\nu = 1/2$ in the mean-field limit. The fit of our results in Fig.~\ref{fig: metrop_m_t_Tc_up_log} to Eq.~\eqref{eqn:power_eqn} yields the estimate $z \approx 2.19$ for $R = 1$ and $z \approx 1.96$ for $R = 128$, which are
consistent with previous results~\cite{z,Lou}. Note that no time scale is associated with the evolution of $m(t)$. 

We next determined $n_s(t)$, the number of clusters of size $s$ at
time $t$ after the temperature quench. Because all the spins are up at $t =
0$, the number of (down) clusters of size $s$ begins at zero and increases
to its (apparent) equilibrium value $n_{s,\,\eq}$. The value of the latter 
depends on the size of the system.

\begin{figure}[t] 
\begin{center}
\scalebox{0.8}{\includegraphics{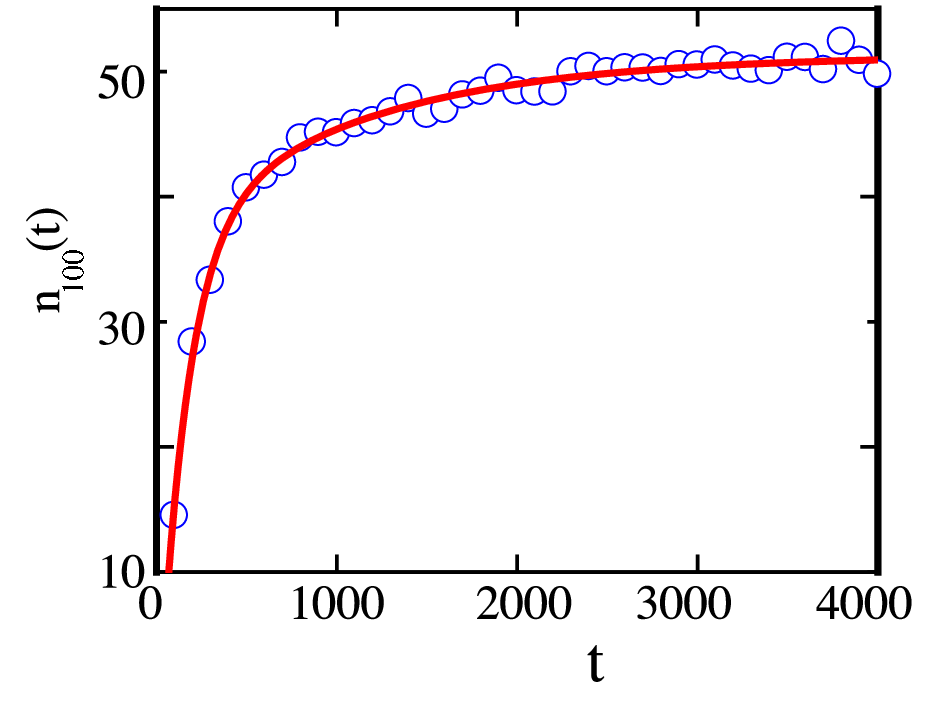}}
\vspace{-0.25cm}
\caption{\label{fig:metrop_ns_100_multiexp_t269_up}The evolution of the
number of clusters of size $s = 100$ at $T = T_{c}$ starting from
$T_0=0$. The fit to Eq,~\eqref{eqn:multi_exp} gives $n_{s,\eq} = 51.3$, $C_{1} = -42$, $C_{2} =
-15$, $\tau_1 = 156$, and $\tau_2 = 1070$.}
\end{center}
\vspace{-0.25cm}
\end{figure}

Figure~\ref{fig:metrop_ns_100_multiexp_t269_up} shows the evolution of
clusters of size $s = 100$ for one run. Because we know of no argument for
the time dependence of $n_s(t) - n_{s,\,\eq}$ except in the mean-field limit~\cite{Lou}, we have to rely on empirical
fits. We find that the time-dependence of $n_s(t)$ can be fitted to the sum of two
exponentials,
\begin{equation}
\label{eqn:multi_exp}
n_{s}(t)-n_{s,\,\eq} = C_{1}e^{-t/\tau_{1}} + C_{2}e^{-t/\tau_{2}},
\end{equation}
where $C_1$, $C_2$, $\tau_{1}$, and $\tau_{2}$ are parameters to be fitted with $\tau_2>\tau_1$.

\begin{figure}[t] 
\begin{center}
\subfigure[\ $T_0=0$.]{\scalebox{0.7}{\label{fig15a}\includegraphics{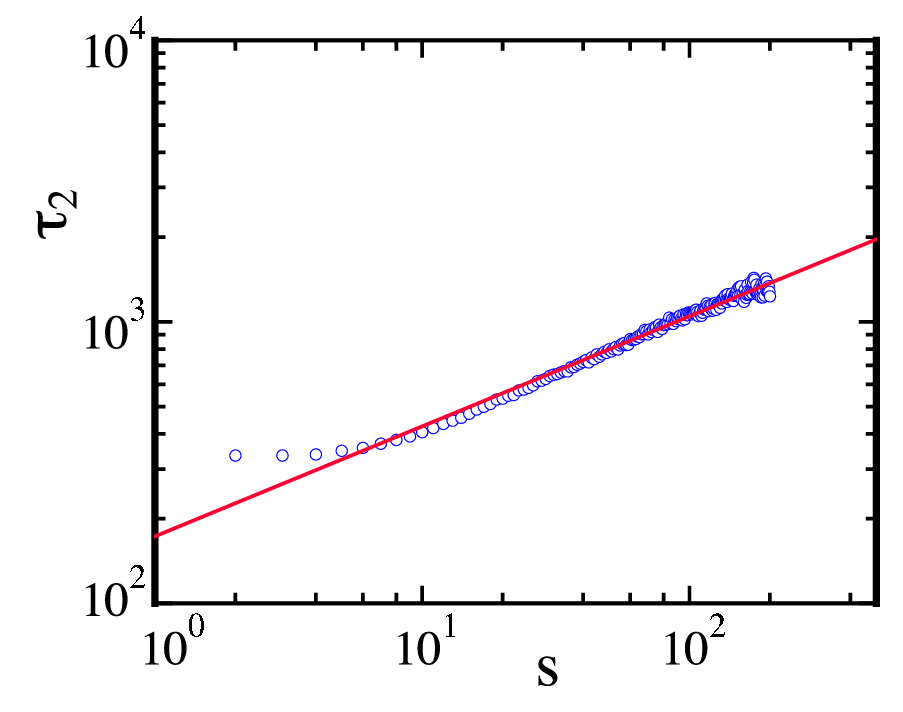}}}
\subfigure[\ $T_0=\infty$.]{\scalebox{0.7}{\label{fig15b}\includegraphics{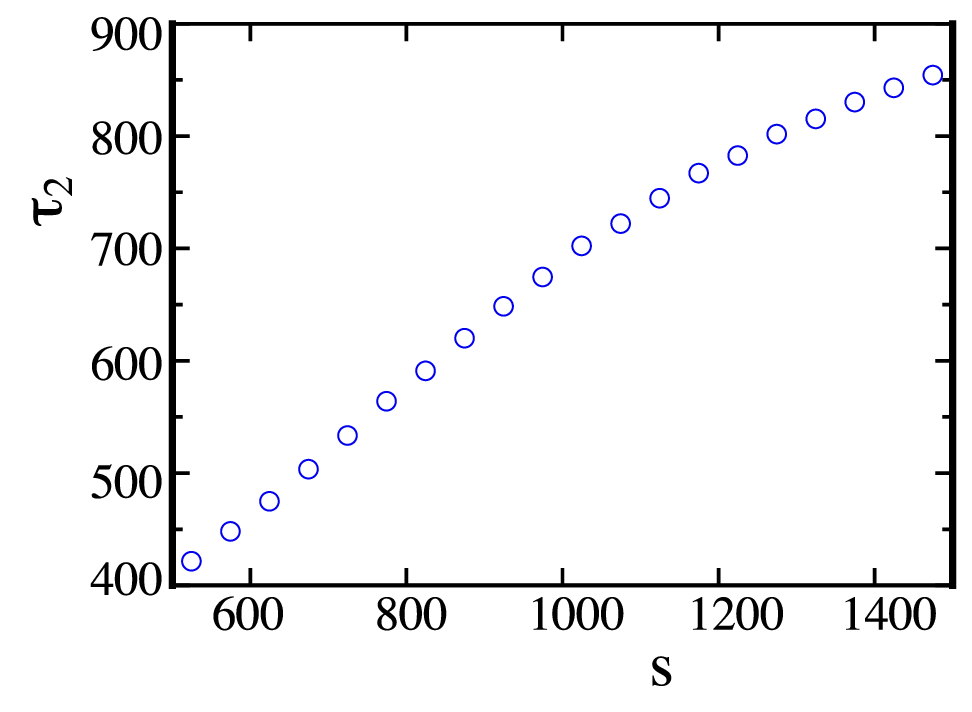}}}
\caption{\label{fig:metrop_tau_s_multiexp_ti_0_tf_269}
The relaxation time $\tau_2$ versus the cluster size $s$ at $T = T_{\rm c}$ for $R = 1$ starting from
(a) $T_{0} = 0$ and (b) $T_{0} = \infty$. The log-log plot in (a) yields $\tau_2 \sim s^{0.4}$.}
\end{center}
\end{figure}

Figure~\ref{fig15a} shows the relaxation
time $\tau_{2}$ as a function of $s$ for $R = 1$ at $T = T_{\rm c}$ starting from $T_{0} = 0$. Note that the bigger the cluster, the longer
it takes to reach its equilibrium distribution. That is, small clusters
form first, and larger clusters are formed by the merging of smaller ones.
The $s$-dependence of $\tau_{2}$ can be approximately fitted 
to a power law with the exponent 0.4.

\begin{figure}[t] 
\begin{center}
\subfigure[\ $s = 30$.]{\scalebox{0.7}{\includegraphics{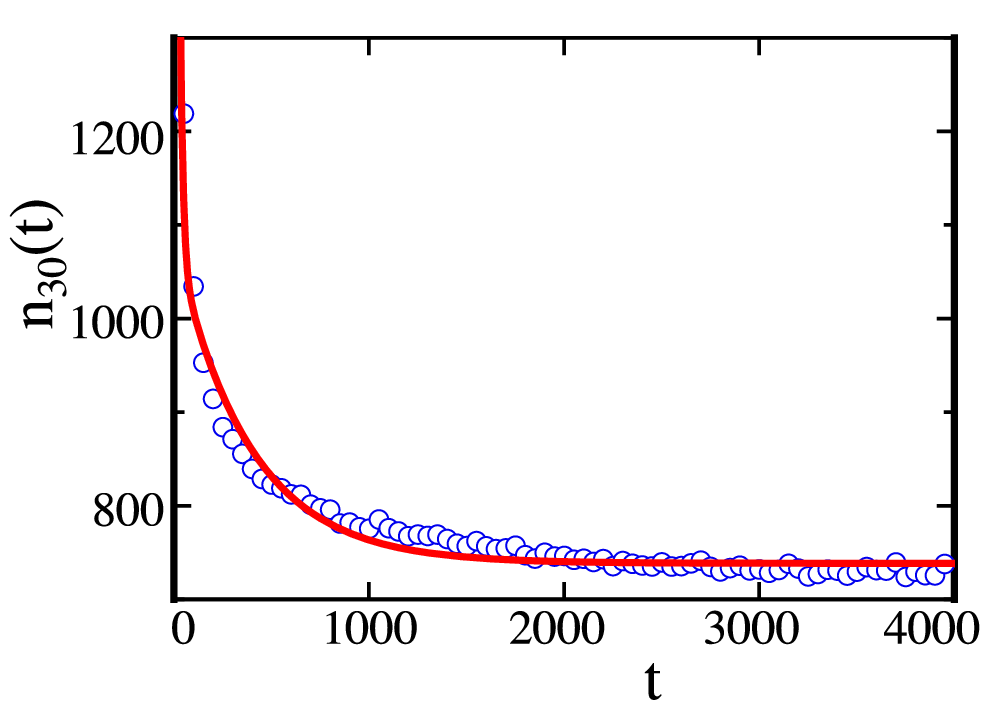}}} 
\subfigure[\ $s = 3000$.]{\scalebox{0.7}{\includegraphics{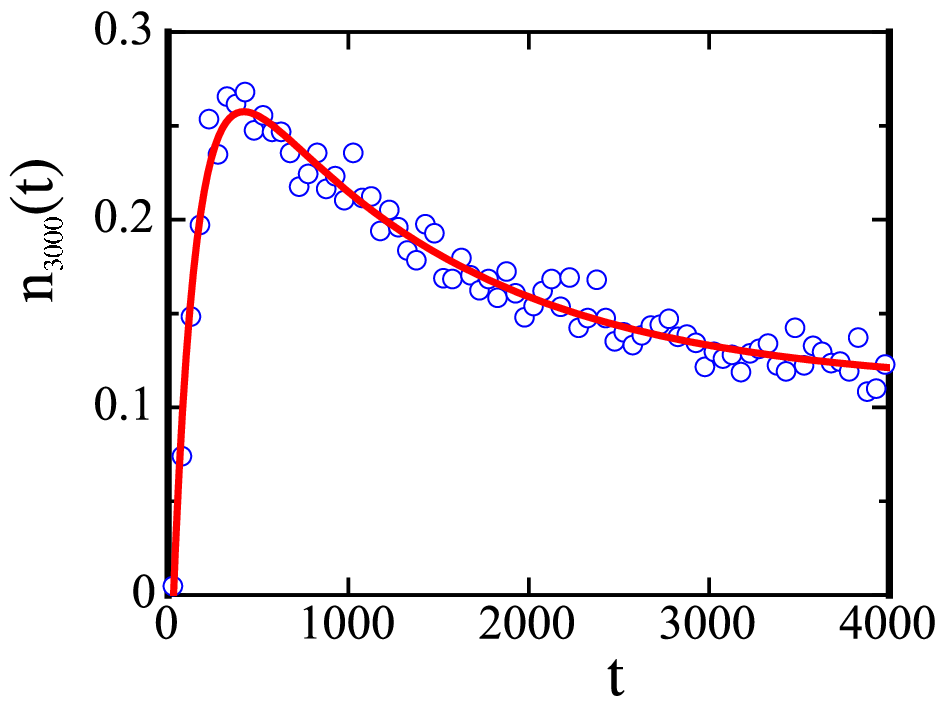}}}
\caption{\label{fig:metrop_t269_inf_down_n_30_vs_t}The time dependence of
the number of clusters of size $s=30$ and $s= 3000$ at $T = T_{\rm c}$ for $R = 1$ starting from $T_0=\infty$. Note that $n_{s= 30}$
monotonically decreases to its equilibrium value and $n_{s=
1000}$ overshoots its equilibrium value. (a) $C_{1} = 2367$, $C_{2} = 332$, $n_{s=30,\,\eq} =
738$,
$\tau_{1} = 16$, and $\tau_{2} = 403$. (b) $C_{1} = -0.42$, $C_{2} =
0.22$, $n_{s=3000,\,\eq} = 0.11$, $\tau_{1} = 130$, and $\tau_{2} = 1290$.}
\vspace{-0.5cm}
\end{center}
\end{figure}

To prepare a configuration at $T_0 = \infty$, the system is randomized with
approximately half of the spins up and half of the spins down. The
temperature is instantaneously changed to $T = T_c$. As before, we 
focus on the relaxation of down spin clusters. In contrast to the $T_0=0$
case, the evolution of the clusters falls into three classes (see
Fig.~\ref{fig:metrop_t269_inf_down_n_30_vs_t}). For small clusters ($1 \leq
s \leq 40$), $n_{s}$ monotonically decreases to its equilibrium value.
This behavior occurs because the initial random configuration has an
abundance of small clusters so that lowering the temperature causes the
small clusters to merge to form bigger ones. For intermediate size clusters
($40 < s < 4000$), $n_{s}$ first increases and then decreases to its
equilibrium value. The initial growth is due to the rapid coalescence of
smaller clusters to form intermediate ones. After there are enough
intermediate clusters, they slowly coalesce to form bigger clusters, which
causes the decrease. For clusters with $s > 4000$, $n_{s}$ slowly
increases to its equilibrium value. The range of sizes for these different
classes of behavior depends on the system size. In all three cases $n_s(t)$
can be fitted to the sum of two exponentials. One of the two
coefficients is negative for $40 < s < 4000$ for which
$n_s(t)$ overshoots its equilibrium value. The relaxation time $\tau_{2}$ is
plotted in Fig.~\ref{fig15b} as a
function of $s$.

\begin{acknowledgments}
We thank Aaron O.\ Schweiger for very useful discussions.
Bill Klein acknowledges the support of Department of Energy grant \# DE-FG02-95ER14498 and Kipton Barros was supported in part by the National Science Foundation grant \# DGE-0221680. Hui Wang was supported in part by NSF grant \# DUE-0442581. The simulations at Clark University were done with the partial 
support of NSF grant \# DBI-0320875.
\end{acknowledgments}


\begin{thebibliography}{33}
\bibitem{moreprecise}More precisely, the
Gibbs assumption is correct in systems for which the interaction range $R \gg 1$ and which are not too close to the spinodal. See Ref.~\onlinecite{bigklein} for more details.
\bibitem{equil}We assume that when the nucleation rate and mean number of clusters of a given size become apparently independent of time that they have reached their equilibrium values.
\bibitem{ising_analytical}The density profile of the \drop\ of the Ising model has been calculated analytically in the limit $R \to \infty$ (K. Barros, unpublished). The result is consistent with the form in Eq.~\eqref{eq:profile} with $A = 3 (\beta J)^{-1} \sqrt{\Delta h/\phi_{\rm s}}$, $w = 3^{-1/2}\phi_{\rm s}^{-1/4} \Delta h^{-1/4}$, and $\rho_0 = -\phi_{\rm s} - A/3$, where $\phi_{\rm s} = \sqrt{1 - (\beta J)^{-1}}$ is the magnitude of $\phi$ at the spinodal. From this analytical solution, the calculated parameters are found to be $A = 0.085$, $w = 2.65$, $\rho_0 = -0.774$ which are very close to the values fitted to the simulation data, $A = 0.084$, $w = 2.45$, $\rho_0 = -0.764$.
\bibitem{explain_critiria}Consider a fluctuation that decays to the metastable phase under noiseless dynamics. To perform the intervention method we make many copies of the configuration and examine the percentage that grow after a given waiting time. Although the expected drift is a decay to the metastable phase, every copy has time to sample a path in configuration space. It is possible that during this waiting time the majority of copies discover and grow toward the stable phase, contradicting the result from the zero-noise criterion. However, for $R \gg 1$ the sampling path will be dominated by the drift term and the two nucleation criteria agree.
\end{thebibliography}
\end{document}